\documentclass{article}
\usepackage{spconf,amsmath,amssymb,graphicx,hyperref}
\usepackage{float}
\usepackage[ruled,vlined]{algorithm2e}
\usepackage{enumitem}
\usepackage{xcolor} 
\usepackage[skip=1pt]{subcaption}
\usepackage[skip=1pt,compatibility=false]{caption}

\DeclareMathOperator*{\argmax}{arg\,max}
\DeclareMathOperator*{\argmin}{arg\,min}

\title{Bayesian Matrix Completion Under Geometric Constraints}
\name{Rohit Varma \qquad Santosh Nannuru}
\address{Signal Processing and Communication Research Center, IIIT Hyderabad}

\begin{document}
\ninept

\maketitle

\begin{abstract}
The completion of a Euclidean distance matrix (EDM) from sparse and noisy observations is a fundamental challenge in signal processing, with applications in sensor network localization, acoustic room reconstruction, molecular conformation, and manifold learning. Traditional approaches, such as rank-constrained optimization and semidefinite programming, enforce geometric constraints but often struggle under sparse or noisy conditions. This paper introduces a hierarchical Bayesian framework that places structured priors directly on the latent point set generating the EDM, naturally embedding geometric constraints. By incorporating a hierarchical prior on latent point set, the model enables automatic regularization and robust noise handling. Posterior inference is performed using a Metropolis-Hastings within Gibbs sampler to handle coupled latent point posterior. Experiments on synthetic data demonstrate improved reconstruction accuracy compared to deterministic baselines in sparse regimes.
\end{abstract}

\begin{keywords}
Euclidean Distance Matrix, Matrix Completion, Bayesian Inference, Distance Geometry, MCMC Methods
\end{keywords}

\section{Introduction}

Matrix completion, the task of inferring missing entries in a partially observed matrix, is a foundational problem in signal processing and machine learning, with applications spanning collaborative filtering, image inpainting, and sensor data reconstruction~\cite{alfakih1999solving}. A particularly challenging subclass involves matrices governed by geometric constraints, such as Euclidean distance matrices (EDMs), which encode squared pairwise distances between points in a low-dimensional space~\cite{fang2012euclidean}. EDMs arise naturally in signal processing contexts, including sensor network localization (SNL), where noisy distance measurements from anchors enable position estimation~\cite{MC_iot}, and acoustic signal analysis, such as reconstructing room shapes from echo delays~\cite{dokmanic2013acoustic}. In molecular biology and crystallography, EDMs facilitate 3D structure determination from inter-atomic distances~\cite{havel1985evaluation}. These measurements are frequently noisy and incomplete, motivating the need for robust denoising and completion methods~\cite{natva2024denoising} 

In general, imputing the missing entries of an arbitrary matrix is a highly ill-posed problem. Without additional structure, there may be infinitely many completions consistent with the observed values. To make the problem tractable, one typically models the problem with structural assumptions such as low-rank, positive semidefiniteness, or geometric constraints to reduce the solution space. 
Low-rank matrix completion methods achieve this by factorizing the target matrix into latent matrices~\cite{jain2013low,cai2010singular}. In probabilistic settings, priors are imposed on these factors~\cite{pmf,bpmf} or, alternatively, directly on rows and columns to promote low-rank structure~\cite{yang2018fast, alquier2014bayesian, paisley2010nonparametric}.

In the case of EDMs, the problem becomes better posed because they obey well-defined properties: their rank is at most $d+2$ for points embedded in $\mathbb{R}^{d}$, 
and although EDMs themselves are not positive semidefinite(PSD), their double-centered form yields a Gram matrix that is PSD. These constraints significantly narrow the space of admissible completions and make robust recovery possible~\cite{edm_survey, exact_recon_abiy_tasissa, lai2017solving, trosset1998applications, oh2001bayesian, keshavan2010matrix}.

Although some traditional methods enforce geometric constraints through objective functions~\cite{parhizkar2013euclidean}, they often struggle with sparse and noisy data due to inflexible, manual regularization. Furthermore, by producing only a {single estimate of the point set}, they fail to quantify the uncertainty~\cite{oh2001bayesian} in the recovered distances, a critical need in many scientific and signal processing applications~\cite{lian2025bayesian,rompelberg2025bayesian}. Our work addresses these gaps by adopting a fully probabilistic inference framework.




We propose a method that mimics the EDM generation process from a point set, ensuring geometric validity. 
We place hierarchical priors on the point set which is nonlinearly related to the EDM.
The specific contributions of this paper are as follows,
\begin{itemize}
\item A Bayesian formulation of EDM completion that incorporates Euclidean geometric constraints, establishing a probabilistic link to nuclear norm minimization for EDMs.
\item A hierarchical model with Normal-Wishart prior and a tailored Metropolis--Hastings within Gibbs sampler, enabling robust inference with adaptive regularization and uncertainty quantification.
\end{itemize}

\section{Problem Formulation}

A matrix $\mathcal{D} \in \mathbb{R}^{n \times n}$ is formally defined as a Euclidean Distance Matrix if there exists a set of $n$ points, represented by the rows of a matrix $P \in \mathbb{R}^{n \times d}$, such that the entries of $\mathcal{D}$ are the squared Euclidean distances between these points,
\begin{align}
\mathcal{D}_{ij} &= \|P_i - P_j\|^2 = \mathcal{D}_{ji} \,,
\end{align}
where $P_i$ and $P_j$ are the $i$-th and $j$-th rows of $P$, respectively, and $d$ is the embedding dimension.
%
The Gram matrix is defined as {$G \in \mathbb{R}^{n \times n} \text{ where } G = PP^T$}. The distance matrix and the Gram matrix are related~\cite{edm_survey} as,
\begin{align}
\mathcal{D}_{ij}  &= \|P_i - P_j\|^2 = G_{ii} - 2G_{ij} + G_{jj} \,, \label{eq:square_distance}\\
{\text{i.e., } \mathcal{D}} &{ = f(G) = \text{diag}(G)\mathbf{1}^T - 2G + \mathbf{1}\text{diag}(G)^T \,,} \label{eq:gram_matrix}
\end{align}
where $\text{diag}(G)$ is a column vector containing the diagonal elements of $G$ and $\mathbf{1}$ is a vector of ones. This equation forms the bridge between the latent point configuration $P$ and the observable distance matrix $\mathcal{D}$. {The linear mapping $f$ from the cone of Gram matrices to Euclidean distance matrices defined by Eq.~\eqref{eq:gram_matrix} is surjective but not injective, as the distance matrix is invariant to translations of the latent points (which alter $G$)~\cite{zhang2016distance}.} Finally, it is easily verified that the ranks of the matrices $G$ and $\mathcal{D}$ are at most $d$ and $d+2$, respectively.

Our observation model assumes that only a subset of distances are seen, corrupted by {additive white Gaussian noise}. Let $\Omega \in \{0,1\}^{n \times n}$ be a binary mask indicating the observed entries. The observed distance $D_{ij}$ is modeled as,
\begin{align}
D_{ij} = \mathcal{D}_{ij} + \epsilon_{ij}, \quad \forall (i,j) \in \Omega \,,
\end{align}
where $\epsilon_{ij} \sim \mathcal{N}(0, \sigma^2) \text{ and }  \sigma^2$ is the noise variance. Thus, the likelihood of observing the distance matrix $D$ given the latent points $P$ is,
\begin{align}
    p(D | P, \sigma^2, \Omega) &= \prod_{i<j} \mathcal{N} \Big(D_{ij} \Big| \|P_i - P_j\|^2, \sigma^2\Big)^{\Omega_{ij}} \,.
    \label{eq:likelihood}
\end{align}
The objective is to estimate a symmetric matrix $\widehat{D}$ that satisfies the geometric constraints of a valid EDM given the set of incomplete and noisy observations $\{D_{ij}\}_{(i,j) \in \Omega}$ .
    
\section{Background}

\subsection{Nuclear norm minimization}
Low-rank matrix completion can be posed as a rank minimization problem. However, this formulation is known to be NP-hard~\cite{candes2012exact}. A common approach is to replace the rank function with its convex surrogate, the nuclear norm, which yields a tractable convex optimization problem that recovers the minimum-rank solution~\cite{recht2010guaranteed}. For the specific case of EDM, ~\cite{lai2017solving, zhang2016distance} reformulate the problem {via this convex relation} as a nuclear norm minimization of the Gram matrix, which is shown to be equivalent~\cite{exact_recon_abiy_tasissa} to,
\begin{align}
    \argmin_P \; \text{Trace}(PP^T) + \frac{\lambda}{2} \sum_{i<j} \Omega_{ij}(D_{ij}-\|P_i - P_j\|^2)^2 \,.
    \label{nnm}
\end{align}
%
This {factorized formulation} provides a tractable approach to low-rank EDM completion and has been shown to succeed under sparse sampling assumptions. However, it requires manual selection of regularization parameter $\lambda$. 


\subsection{Probabilistic framework}
Probabilistic factorization approaches model matrix completion as estimation of their low rank latent matrices. The foundational model is probabilistic matrix factorization (PMF)~\cite{pmf}, which models a partially observed matrix $R^{n \times m}$ using low-dimensional latent factors $U^{d \times n}$ and $V^{d \times m}$ having Gaussian priors. The ${ij}$-th matrix entry $R_{ij}$ is modeled as,
\begin{align}
& p(R_{ij} | U_i, V_j, \sigma^2) = \mathcal{N}(R_{ij} | U_i^T V_j, \sigma^2) \,, \label{eq:ratings_likelihood}\\
& p(U_i | \sigma_u^2) = \prod_{i=1}^{n} \mathcal{N}(U_i | 0, {\sigma_u}^2) \,, \;
p(V_i | \sigma_v^2) = \prod_{i=1}^{m} \mathcal{N}(V_i | 0, {\sigma_v}^2) \,. \nonumber
\end{align}
%
The latent factors are estimated by finding the maximum a posteriori (MAP) estimate from the log posterior. This method is highly hyperparameter dependent and overfitting can be a severe issue in sparse settings. 

The work in~\cite{bpmf} extends PMF to a fully Bayesian setting. The Bayesian PMF (BPMF) assumes slightly different Gaussian priors on the latent factors,
\begin{align}
U_i \sim \mathcal{N}&(\mu_U, \Lambda_U^{-1}), \qquad 
V_j \sim \mathcal{N}(\mu_V, \Lambda_V^{-1}) \,,
\end{align}
%
and additionally models the hyperparameters $(\mu_U, \Lambda_U)$ and $(\mu_V, \Lambda_V)$ with conjugate Gaussian–Wishart distribution. This yields tractable Gibbs updates for all latent variables, allowing efficient posterior sampling without manual regularization.

In the specific case of EDM completion,~\cite{xue2019locating} proposed a Bayesian low-rank factorization method, where the EDM is expressed as the product of two latent matrices $A$ and $B$, placed with hierarchical priors. However, this formulation does not explicitly capture the geometric constraints inherent to EDMs.


A geometrically faithful model for EDM completion encounters a significant challenge: the likelihood in~\eqref{eq:likelihood} depends on squared Euclidean distances~\eqref{eq:square_distance} which is quadratic in $P$. 
In contrast, the likelihood~\eqref{eq:ratings_likelihood} in PMF \cite{pmf} model is bilinear in the latent factors $U$ and $V$ yielding Gaussian conditional distributions. The geometric constraint along with the coupling between latent points complicates the posterior geometry, preventing direct extension of BPMF. 



\section{Bayesian framework for EDM}
We formulate the EDM completion and denoising problem within a Bayesian framework. To do this, we place priors on the latent points $P_i$. The posterior for the EDM is obtained by combining the prior and the likelihood~\eqref{eq:likelihood} using the Bayes rule. Inspired by the probabilistic modeling in~\cite{pmf} and~\cite{bpmf}, we propose two prior structures: (a) Gaussian prior and (b) hierarchical Gaussian prior with conjugate Normal-Wishart hyperprior. The priors help in regularizing the model and enable robust inference. Maximum a posteriori (MAP) inference is performed when using Gaussian prior and Markov chain Monte Carlo (MCMC)~\cite{robert1999monte} inference when using hierarchical prior.





\subsection{Inference with Gaussian prior}
We place an independent zero-mean Gaussian prior on each point $P_i$,
\begin{equation}
    p(P | \sigma_p^2) = \prod_{i=1}^{n} \mathcal{N}(P_i \Big| \mathbf{0}, \sigma_p^2 I_d)
    \label{gaussian_prior}
\end{equation}
where $\sigma_p^2$ is the variance and $\mathbf{0}$ is the vector of all zeros.
%
We seek a point estimate for $P$ by maximizing the posterior probability (MAP). This is equivalent to minimizing the negative log-posterior, 
\begin{align}
    &P^* = \argmax_{P} \; \log p(D|P,\sigma^2,\Omega) + \log p(P|\sigma_p^2) \\
        &= \argmin_{P} \frac{\lambda}{2} \sum_{i<j} \Omega_{ij}(D_{ij}-\|P_i - P_j\|^2)^2 + \|P\|_F^2 \,.
        \label{eq:map_estimate}
\end{align}
%
This MAP formulation is equivalent to the formulation in~\eqref{nnm} where {$\lambda = 2\sigma^2_p / \sigma^2$}. Thus, for the matrix completion problem, nuclear norm minimization approach is equivalent to the MAP estimate assuming Gaussian prior on the latent points. 
%
This objective function can be solved using the augmented Lagrangian \cite{exact_recon_abiy_tasissa}.

While this MAP framework provides a useful point estimate and establishes a clear link to regularized optimization, it has two main limitations: (1) it does not quantify the uncertainty of the missing entries, and (2) it relies on fixed prior variance~\eqref{eq:map_estimate} which acts as a rigid regularization parameter and needs to be carefully tuned~\cite{bpmf}.

A full Bayesian treatment using MCMC could be applied to this Gaussian model to address the first limitation. However, to overcome the second and more critical limitation, we opt for a more sophisticated hierarchical prior model. 

\subsection{Inference with hierarchical Gaussian prior}

The hierarchical prior enables the model to learn the level of regularization from the data, since the associated hyperparameters are integrated out in Bayesian inference.

\subsubsection{Hierarchical Prior}
We introduce a flexible prior on the latent points $P$ by assuming they are drawn from a Gaussian distribution with an unknown mean $u_p$ and an unknown precision matrix $\Delta_p$. To complete the hierarchy, we place a conjugate Normal-Wishart hyperprior on the mean and precision matrix,
\vspace{-0.4cm}
\begin{align}
    p(P | u_p, \Delta_p) &= \prod_{i=1}^n \mathcal{N}(P_i | u_p, \Delta_p^{-1}) \,,\label{eq:P_prior} \\
    p(u_p, \Delta_p) &= \mathcal{N}(u_p | u_0, (\beta_0 \Delta_p)^{-1}) \mathcal{W}(\Delta_p | W_0, \nu_0) \,, \label{eq:hyperprior}
\end{align}
where $\Theta_0 = \{u_0, \beta_0, W_0, \nu_0\}$ are fixed hyperparameters of the model. This prior structure is analogous to the one used in~\cite{bpmf}.

\subsubsection{Posterior Inference}

%

Due to the non-linear relationship between the latent points $P$ and the observed distances $D$, the joint posterior distribution over model parameters $\{P, u_p, \Delta, \sigma^2\}$ is analytically intractable. Variational methods approximate the posterior using factors of simple distributions, which could potentially give inaccurate results. To perform inference, we therefore turn to 
Markov chain Monte Carlo (MCMC), which is asymptotically exact~\cite{robert1999monte}. We use Gibbs sampling to efficiently draw samples from the conditional posteriors of the hyperparameters, a step made possible by our choice of conjugate priors. However, the conditional posterior for the latent points $P$ is non-standard and cannot be sampled directly. Therefore, we use Metropolis-Hastings (MH) for sampling $P$.\\
\noindent\textbf{Hyperparameter Sampling:}
Given the current state of the latent points $P^{(t)}$, we can sample the hyperparameters $(u_p, \Delta_p)$ exactly due to the conjugacy of the Normal-Wishart hyperprior, meaning the conditional posterior for the hyperparameters also takes the form of a Normal-Wishart distribution.

To derive the parameters of this posterior, we combine the likelihood of the points~\eqref{eq:P_prior} with the Normal-Wishart prior~\eqref{eq:hyperprior}. After simplification, the resulting posterior parameters ${\Theta_p =} (\beta^*, u^*, \nu^*,$ $W^*)$ can 
be expressed as updates based on the empirical statistics of $P^{(t)}$. We can then sample directly from this updated distribution.



\vspace{-0.4cm}
\begin{align}
    &p(u_p, \Delta_p | P) = \mathcal{N}(u_p | u*, (\beta^* \Delta_p)^{-1}) \mathcal{W}(\Delta_p | W*, \nu*) \,, \label{eq:hyperparameter_posterior}\\
    &\bar{P} = \frac{1}{n}\sum_{i=1}^n P_i, \qquad \bar{S} = \sum_{i=1}^n (P_i - \bar{P})(P_i - \bar{P})^\top \label{eq:theta_p_interim_update}\\
    &\nonumber \beta^* = \beta_0 + n \qquad
    u^* = \frac{\beta_0 u_0 + n\bar{P}}{\beta_0 + n} \qquad \nu^* = \nu_0 + n \\  
    &(W^*)^{-1} = W_0^{-1} + S + \frac{\beta_0 n}{\beta_0 + n}(\bar{P} - u_0)(\bar{P} - u_0)^\top 
    \label{eq:theta_p_update}
\end{align}

\noindent\textbf{Latent Point Sampling:}  The conditional posterior for each latent point $P_i$ can be written as follows 
\begin{align}
&p\big(P_i\mid  P_{-i}, \mathcal{D}, \Theta_p, \sigma^2)\big) 
\propto \\
& \nonumber \exp\!\Big(\!
- \tfrac{1}{2}(P_i-u_p)^\top\Delta_p(P_i-u_p) - \frac{1}{2\sigma^{2}}\sum_{(i,j) \in \Omega}r_{ij}^{2}\!
\Big)
\label{eq:p_posterior}
\end{align}
where the residual is $r_{ij} = D_{ij} - ||P_i - P_j||^2$ {and $P_{-i} = \{\, P_j \mid j \in \{1,\dots,n\},\; j \neq i \,\}$.} This distribution is non-Gaussian due to

\begin{algorithm}[t]
\caption{MCMC Sampling Iteration}

\KwIn{Current state $\{P^{(t)}, u_p^{(t)}, \Delta_p^{(t)}, \alpha^{(t)}\}$} 
\KwOut{Next state $\{P^{(t+1)}, u_p^{(t+1)}, \Delta_p^{(t+1)}, \alpha^{(t+1)}\}$}
\BlankLine

\textbf{Hyperparameter Update (Gibbs):}\\
{Compute $\bar{P}^{(t)}$ and $\bar{S}^{(t)}$ using \eqref{eq:theta_p_interim_update}.\\
Update  $\Theta_p = (\beta^*, u^*, \nu^*, W^*)$ using \eqref{eq:theta_p_update}}.\\
Sample $\Delta_p^{(t+1)} \sim \mathcal{W}(W^*, \nu^*)$.\\
Sample $u_p^{(t+1)} \mid \Delta_p^{(t+1)} \sim \mathcal{N}(u^*, (\beta^* \Delta_p^{(t+1)})^{-1})$.\\

\textbf{Latent points update (Metropolis–Hastings):}\\
\For{$i = 1, \dots, n$}{
    Propose $P_i^* \sim \mathcal{N}(P_i^{(t)}, \tau^2 I_d)$, where $\tau$ is the proposal std dev.\\
    With probability $\rho$, set $P_i^{(t+1)} \gets P_i^*$; else $P_i^{(t+1)} \gets P_i^{(t)}$.} 
\textbf{EDM estimate:}\\
\vspace{-6pt}
\begin{equation}
\hat{D_{ij}} = ||P^{(t+1)}_i - P^{(t+1)}_j||^2 \;\; \forall \, i,j
\label{eq:EDM_Update}
\end{equation}

\textbf{Precision update (Gibbs):}\\
Update $a^*, b^*$. \\
Sample $\alpha^{(t+1)} \sim \text{Gamma}(a^*, b^*)$
\label{alg:mcmc}
\end{algorithm}

\noindent the term $\|P_i - P_j\|^2$. We therefore use a Metropolis-Hastings step to sample each $P_i$. For each point, we propose a new position and accept it based on an acceptance probability that depends on the likelihood and the prior. The acceptance probability $\rho$ is,
\begin{align}
     \rho = \min\left(1, \frac{p(P_i^* | P_{-i}, \mathcal{D}, \Theta_p, \sigma^2)}{p(P_i^{(t)} | P_{-i}, \mathcal{D}, \Theta_p, \sigma^2)}\right) \,.\label{eq:proposal_prob}
\end{align}
\noindent where $P^*_i$ is the proposal at step $t$.\\
\noindent \textbf{Precision update:} Let $\alpha = \sigma^{-2}$ denote the noise precision. The model learns the precision to adapt to different noisy setups  and is not bound by the fixed assumption about noise level. The conditional posterior combines the likelihood with a Gamma prior
$\text{Gamma}(a_0, b_0)$:
\begin{align}
& p(\alpha \mid D_{\Omega}, P) \propto p(D_{\Omega} \mid P, \alpha)\, p(\alpha) \nonumber \\
&\propto 
   \alpha^{\,a_0 + \tfrac{|\Omega|}{2} - 1} 
   \exp\!\Bigg(
      -\alpha \Bigg[
         b_0 + \tfrac{1}{2}\!\sum_{(i,j)\in\Omega} r_{ij}^2
      \Bigg]
   \Bigg). \\
\alpha &\sim \text{Gamma}(a^*, b^*), \nonumber \\
a^* &= a_0 + \tfrac{|\Omega|}{2},  
\qquad
b^* = b_0 + \tfrac{1}{2}\!\sum_{(i,j)\in\Omega} r_{ij}^2.
\end{align}


This hybrid sampling scheme in Algorithm \ref{alg:mcmc} allows us to efficiently explore the posterior distribution, providing not only estimates for the missing entries but also a full quantification of their uncertainty.







\vspace{-0.4cm}
\section{Simulation Results}
\vspace{-0.1cm}

We evaluate the proposed Bayesian matrix completion under geometric constraints (BMC-GC) algorithm using simulations. 
The source code~\cite{code} for our algorithm is publicly available. For comparison, we have used the open-source implementations of the baseline methods~\cite{exact_recon_abiy_tasissa,edm_survey}.
For each experiment, we generate a ground-truth point set $P\in \mathbb{R}^{n\times d}$ by drawing $n$ points from a $d$-dimensional standard normal distribution. We simulate points in 3-dimensional space, setting the embedding dimension to $d = 3$. 

\begin{figure*}[t]
    \centering
    \begin{subfigure}[t]{0.31\linewidth}
        \centering
        \includegraphics[width=\linewidth]{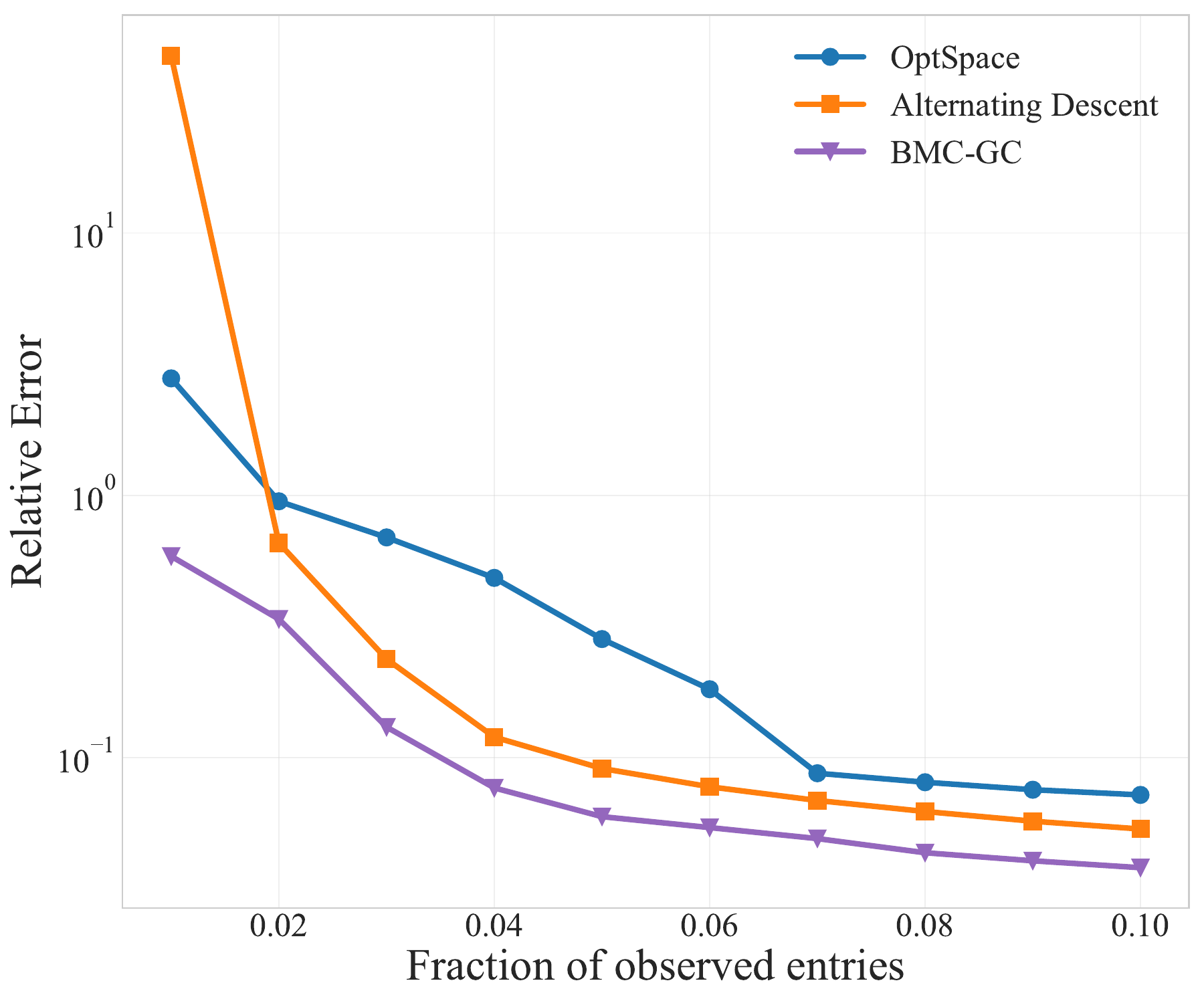}
        \caption{{Relative Error vs Fraction of observed entries}} \label{fig:noise_ablation_a}
    \end{subfigure}\hfill
    \begin{subfigure}[t]{0.34\linewidth}
        \centering
        \includegraphics[width=\linewidth]{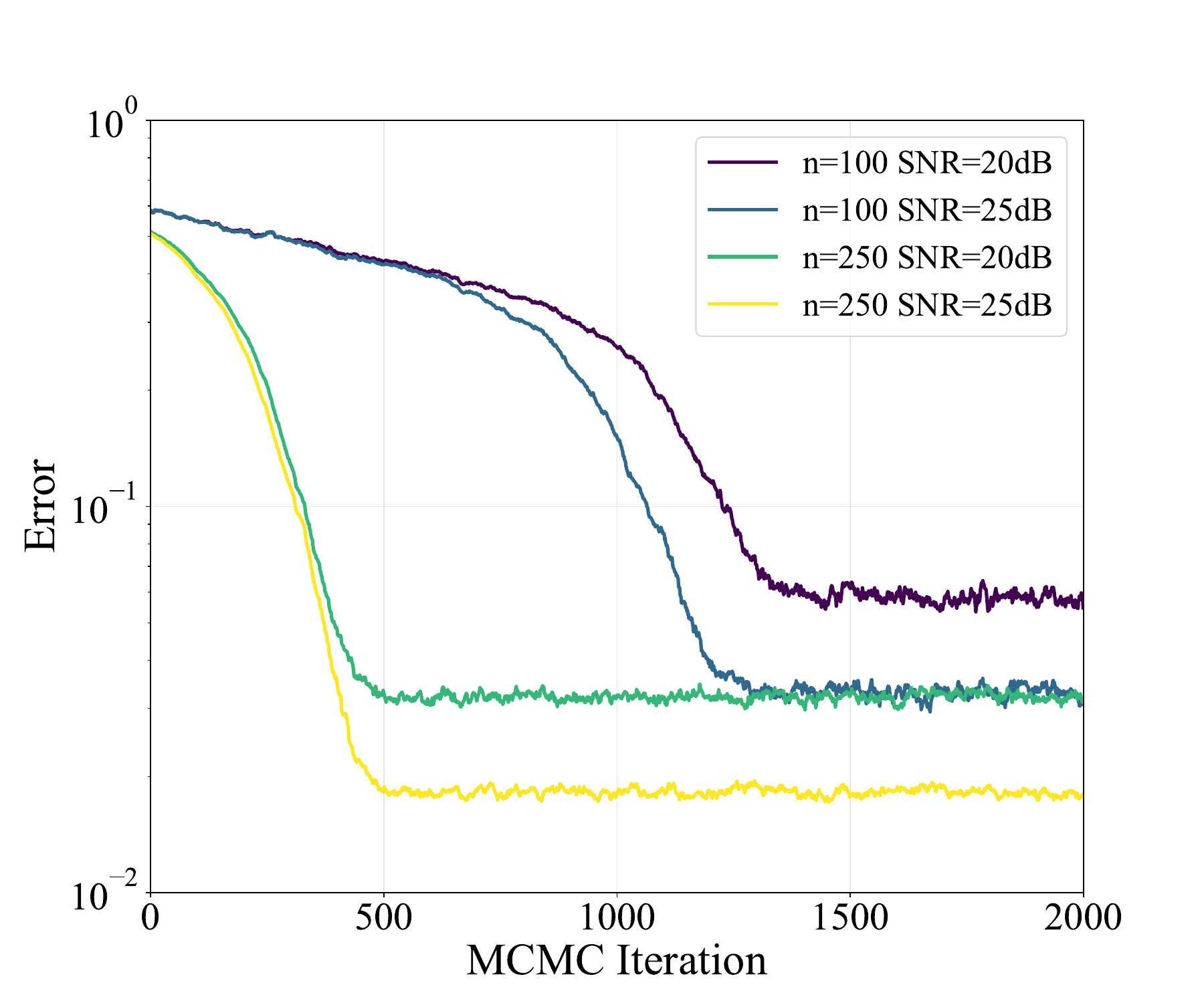}
        \caption{Relative Error convergence} \label{fig:noise_ablation_b}
    \end{subfigure}\hfill
    \begin{subfigure}[t]{0.34\linewidth}
        \centering
        \includegraphics[width=\linewidth]{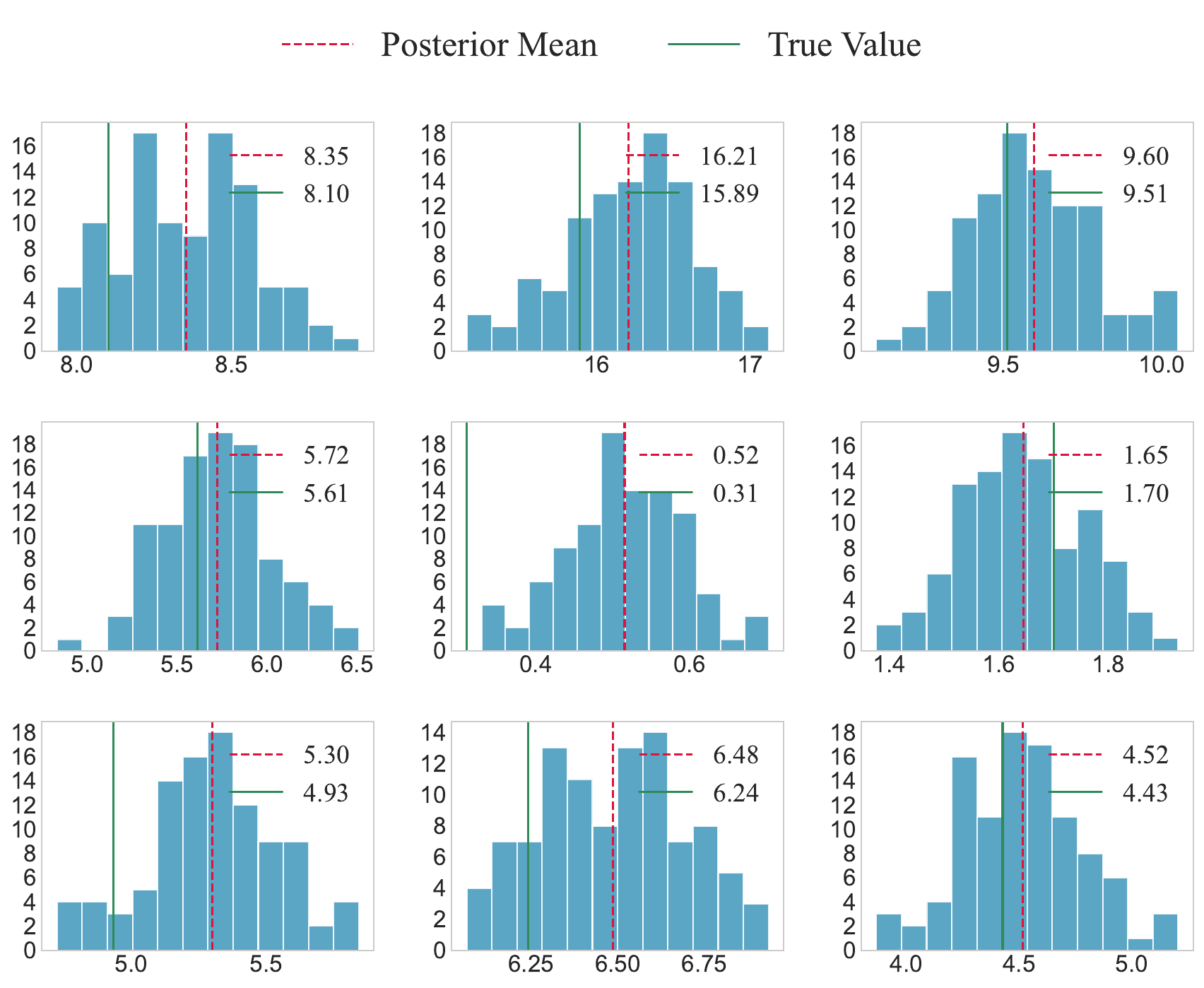}
        \caption{Uncertainty Quantification of estimated EDM} \label{fig:noise_ablation_c}
    \end{subfigure}
    \caption{Performance evaluation of BMC-GC on EDM denoising and completion. (a) Error vs fraction of observed entries under noisy conditions ($n=500$, SNR = 20 dB). (b) Convergence of the proposed sampler, showing relative reconstruction error over MCMC iterations with fraction of observed entries $0.2$. (c) 
    {Posterior distribution of 9 randomly selected missing entries of the EDM. Each subplot displays the posterior distribution of samples for a distinct unobserved pair $(i,j)$, along with the True Value (red) and Posterior Mean (blue). Results shown for $n=100$, fraction of observed entries $=0.5$, and $SNR=20$ dB.}}
    \label{fig:convergence_ablation}
\end{figure*}

\noindent entries of the true EDM are $\mathcal{D}_{ij} = ||P_i - P_j||^2$. To simulate partial observations, we generate a symmetric binary mask(uniformly random) $M$ indicating observed entries. For experiments involving noise, we add independent zero-mean Gaussian noise to the observed entries of $\mathcal{D}$. If $N$ is the noise matrix, the signal-to-noise ratio is defined as $\text{SNR} = 10 \log_{10} \frac{|| \mathcal{D} \odot M ||_F^2}{|| N \odot M ||_F^2}$, where the Hadamard product $\odot$ denotes element-wise multiplication.

The results are obtained using the proposed Metropolis-within-Gibbs sampler over 20 independent trials. Each trial generates a randomized $\mathcal{D}$, $M$, and $N$.
For each trial, we run the MCMC chain for a total of 1500 iterations, {we choose a conservative burn-in~\cite{robert1999monte} of 1200 to ensure convergence across a range of simulation parameters, specifically for smaller $n$ and sparse observations.}
The final EDM estimate $\widehat{D}$ is the average of 30 posterior samples~\eqref{eq:EDM_Update}, collected by thinning the chain every 10 iterations after burn-in. The performance of the algorithm is quantified by the relative reconstruction error measured as, 
\begin{equation}
E = \frac{|| \widehat{D} - \mathcal{D} ||_F}{|| \mathcal{D} ||_F} \,.
\end{equation}

We compare the proposed method with the following methods from literature: OptSpace \cite{keshavan2010matrix}, Alternating descent \cite{parhizkar2013euclidean}, and augmented Lagrangian \cite{exact_recon_abiy_tasissa}. The hyperparameters used for the experiments are 
$\tau = 0.05, \; a_0 = 10^{-6}, \; b_0 = 10^{-6}, \; 
\beta_0 = 2, \; \nu_0 = d+2, \; u_0 = \mathbf{0}^{d\times1}, \; W_0 = I^{d\times d}$.

\subsection{Results}
\begin{figure}[h]
    \centering
    \includegraphics[width=0.7\linewidth]{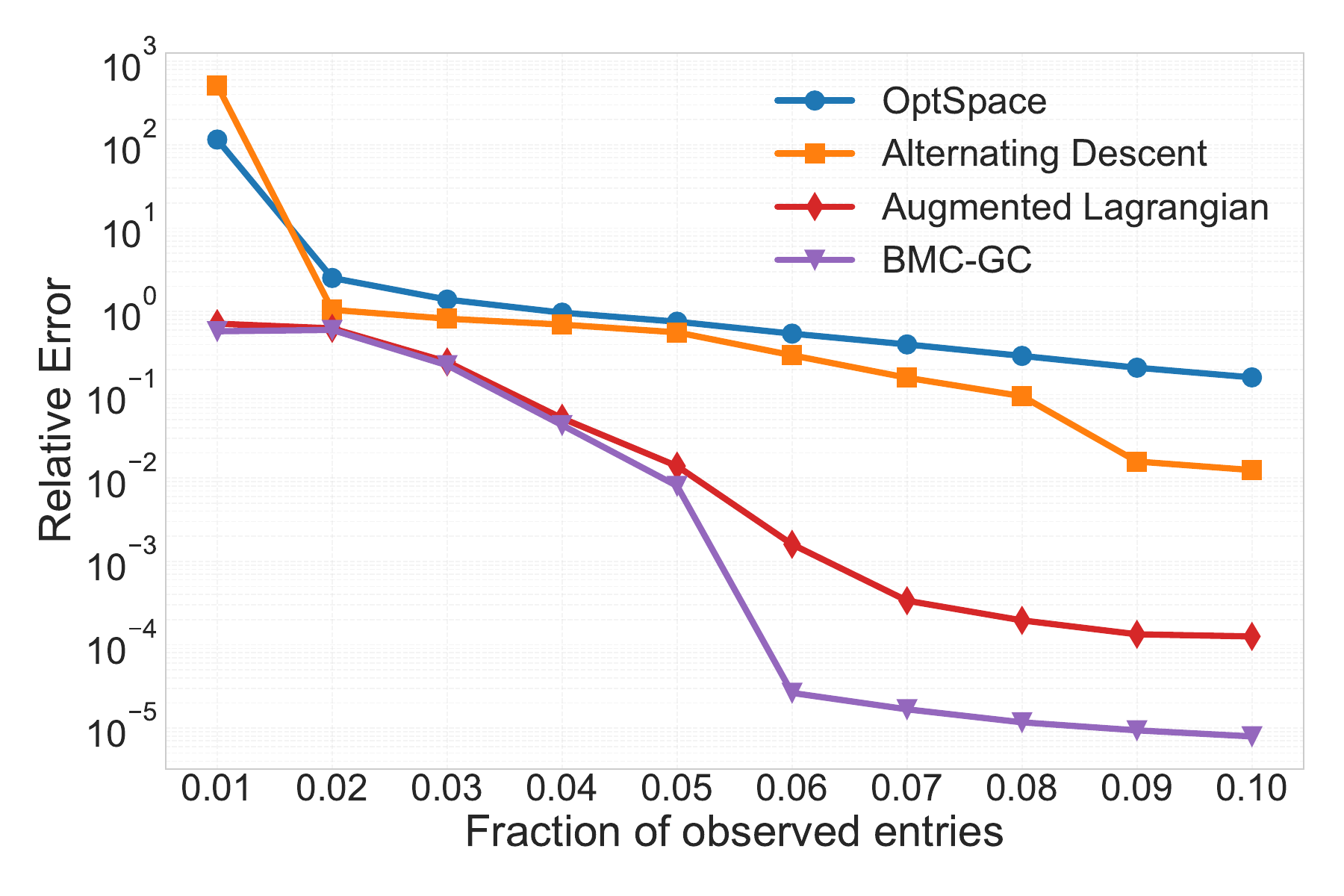}
        \caption{Reconstruction Error in noiseless setup with $n=250$}
    \label{BMC-GC_noiseless250}
\end{figure}

\begin{figure}[H]
    \centering
    \includegraphics[width=0.7\linewidth]{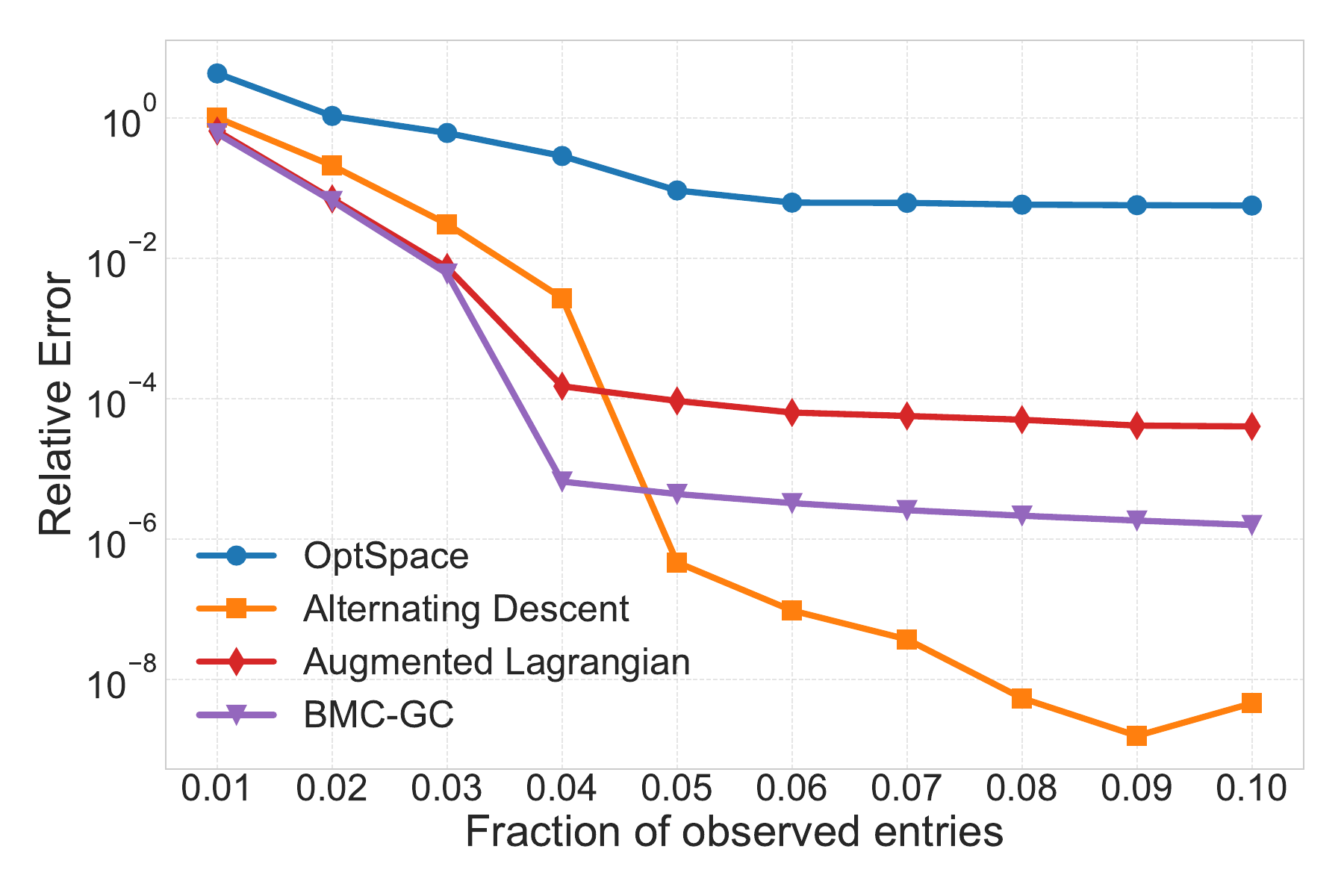}
        \caption{Reconstruction Error in noiseless setup with $n=500$}
    \label{BMC-GC_noiseless500}
\end{figure}

Figure~\ref{fig:noise_ablation_a} demonstrates performance of BMC-GC for varying sparsity under noisy conditions. It outperforms deterministic baselines of OptSpace~\cite{keshavan2010matrix} and alternating descent~\cite{parhizkar2013euclidean}. This highlights the benefit of adaptive regularization which is enabled by hierarchical priors on both the latent points $P$ and the noise precision $\alpha$.
Figure~\ref{fig:noise_ablation_b} shows convergence of error with MCMC iterations. Convergence happens earlier in the presence of larger number of available entries (large $n$) but the error at convergence depends on the SNR.
Figure~\ref{fig:noise_ablation_c} plots histograms of the posterior samples obtained after the burn-in period. The true EDM values are indicated highlighting BMC-GC's ability to quantify reconstruction uncertainty.
In the noiseless case (Figure~\ref{BMC-GC_noiseless250}), BMC-GC demonstrates superior performance in situations with high sparsity, which is particularly evident for the 250 point configuration. 
However, as the number of observed entries increases (Figure~\ref{BMC-GC_noiseless500}), deterministic methods become more effective.

\vspace{-1cm}


\section{Conclusion}
In this paper, we presented a hierarchical Bayesian framework to complete Euclidean distance matrices with sparse and noisy observations. 
We placed a hierarchical prior on the latent points generating the EDM and performed posterior inference using a Gibbs sampler. The Gibbs sampler, in turn, utilized the Metropolis-Hastings algorithm 
to sample from the conditional distribution of the latent points. 
Our approach ensures geometric validity and provides a full posterior distribution of the completed matrix. Simulation results demonstrate the effectiveness of 
the algorithm at high missing rates.

\bibliographystyle{IEEEbib}
\bibliography{strings,refs}

\end{document}